\documentclass[10pt,onecolumn,preprintnumbers,amsmath,amssymb,nofootinbib,superscriptaddress,aps,prd,numberedappendix,a4wide]{openjournal}

\usepackage{newtxtext,newtxmath}
\usepackage{lipsum}
\usepackage[T1]{fontenc}
\usepackage{ae,aecompl}
\usepackage[nodisplayskipstretch]{setspace}

\usepackage{graphicx}	
\usepackage{amsmath}	
\usepackage{amssymb}	

\begin{document}
\vspace{-0.5cm}
\title{Spatial Propagation of Weak Lensing Shear Response Corrections}

\author{T. D. Kitching$^{1}$, N. Tessore$^{2}$, P. L. Taylor$^{3,4,5}$}
\affiliation{
$^{1}$Mullard Space Science Laboratory, University College London, Holmbury St Mary, Dorking, Surrey RH5 6NT, UK\\
$^{2}$Department of Physics \& Astronomy, University College London, Gower Street, London WC1E 6BT, UK\\
$^{3}$Center for Cosmology and AstroParticle Physics (CCAPP), The Ohio State University, Columbus, OH
43210, USA\\
$^{4}$Department of Physics, The Ohio State University, Columbus, OH 43210, USA\\
$^{5}$Department of Astronomy, The Ohio State University, Columbus, OH 43210, USA
}

\begin{abstract}
In this paper we show how response function corrections to shear measurements (e.g. as required by Metacalibration) propagate into cosmic shear power spectra. We investigate a 2-sphere pixel (also known as `HEALpixels') correction and a forward-modelling approach using simple Gaussian simulations. In the 2-sphere pixel-correction approach we find a free parameter that is the tolerated condition number of the local response matrices: if this is too large then this can cause an amplification of the shot noise power spectrum, if too small it can lead to a loss of area (and a possible selection bias). In contrast by forward-modelling the power spectrum this choice can be avoided. This also applies to  map-based inference methods using shear-response calibrated maps.
\end{abstract}

\maketitle

\section{Introduction}
\label{S:Intro}
Weak lensing is the effect whereby the apparent observed ellipticity (third flattening, or third eccentricity) of galaxies is altered by the presence of matter along the line-of-sight. The effect can be approximated by an additional ellipticity added to the unlensed (intrinsic) ellipticity that is known as shear. Measurement of the weak lensing effect from data, to infer the shear, can be biased by several effects e.g. inaccuracies in the algorithms used \citep{step1,step2,great08,great10,great3}, detector effects \citep{2014JInst...9C3048A}, the size of the point spread function \citep{2017MNRAS.468.3295H,2019A&A...624A..92K, 2021MNRAS.504.4312G}, and detection effects \citep{cccp, 2021arXiv210810057H}. \cite{K19,K20} demonstrated how biases, parameterised by multiplicative and additive terms that modify the observed ellipticity (unlensed ellipticity plus shear), affect the observed power spectrum of weak lensing data (known as cosmic shear).

However, in several methods such as lensfit \citep{2007MNRAS.382..315M} and Metacalibration \citep[MetaCal;][]{2017ApJ...841...24S,metacal}, there is the concept of an additional multiplicative response matrix that multiplies only the shear and not the unlensed ellipticity measurement. In particular to account for such a term in MetaCal a local pixel-level correction has been proposed, but the effect of the local noise on these corrections, and alternatively the propagation of the response terms through to the cosmic shear power spectrum has not been shown, and this is what we address in this paper. In Section \ref{S:Method} we present the methodology, in Section \ref{S:Results} we present results of testing on simulations, and in Section \ref{Conclusions} we discuss conclusions.

\section{Method}
\label{S:Method}
The MetaCal \citep{metacal, 2017ApJ...841...24S} correction can be written as a local transformation of the ellipticity field like
\begin{equation}
    [\widetilde e_1+{\rm i}\widetilde e_2]_p=[e_1+{\rm i}e_2+
    (R_{11}\gamma_1+R_{12}\gamma_2)+{\rm i}(R_{21}\gamma_1+R_{22}\gamma_2)]_p,
\end{equation}
where $R_{ij}$ are elements of the $2\times 2$ response matrix $R$, $\gamma_i$ are the true shear, $\widetilde e$ are the observed ellipticities, and $e_i$ are the unlensed uncorrelated intrinsic ellipticity component (i.e. the shot noise term). This is a locally defined transform within a angular pixel $p$ where $i=\{1,2\}$ refer to local distortions relative to a Cartesian tangent plane, where $i=1$ are distortions along the axes and $i=2$ are distortions at $45$ degrees to the axes. Locally, the elements of the response matrix can be related to spin-0 and spin-4 multiplicative biases to the shear where
\begin{eqnarray}
\label{p1}
[\widetilde e_1+{\rm i}\widetilde e_2]_p&=&[e_1+{\rm i}e_2+
    r_0(\gamma_1+{\rm i}\gamma_2)+r_4(\gamma_1-{\rm i}\gamma_2)]_p\nonumber\\
r_0&=&\frac{1}{2}[(R_{11}+R_{22})+{\rm i}(R_{21}-R_{12})]\nonumber\\
r_4&=&\frac{1}{2}[(R_{11}-R_{22})+{\rm i}(R_{21}+R_{12})].
\end{eqnarray}
This can be compared to a global expression on the Celestial sphere in which
\begin{equation}  
  \widetilde {\mathbf e}(\mathbf{\Omega})={\mathbf e}(\mathbf{\Omega})+r_0(\mathbf{\Omega}){\mathbf\gamma}(\mathbf{\Omega})+r_4(\mathbf{\Omega}){\mathbf\gamma^*}(\mathbf{\Omega})
\end{equation}
where ${\mathbf x}(\mathbf{\Omega})=x_E(\mathbf{\Omega})+{\rm i}x_B(\mathbf{\Omega})$, where each field is defined relative to an $E$ and $B$-mode field on the sphere, where $\mathbf{\Omega}$ are angular coordinates. $r_0$ is a spin-0 term and $r_4$ is a spin-4 term, where the total spin-2 nature of all terms is conserved. 
\newpage
\subsection{Power Spectra}
The power spectra estimates for the measured ellipticity can now be computed by taking the correlation of the spherical harmonic coefficients, computed using a spin-weight spherical harmonic transform for a spin-2 field, where 
\begin{eqnarray}
\label{eqsize}
\widetilde C^{GH}_{\ell,ij}&\equiv& \frac{1}{2\ell+1}\sum_m 
\widetilde e^G_{\ell m,i}\widetilde e^{H,*}_{\ell m,j}
\end{eqnarray}
for $G=\{E, B\}$ and $H=\{E, B\}$, where $i$ and $j$ labels for tomographic bins delineating galaxy populations defined by redshift or colour \citep{2019arXiv190106495K}.
We will assume that the true $EB$ and $BE$ power spectra are zero $C^{EB}_{\ell,ij}=C^{BE}_{\ell,ij}=0$, which should be the case in all but the most exotic dark energy models that cause parity-violating modes \citep{2013LRR....16....6A}. Given this assumption, the estimated $EE$ power spectra is given by  
\begin{eqnarray}
\label{full}
\widetilde C^{EE}_{\ell,ij}=\left[{\sum_{\ell'}}
{\mathcal M}^{++}_{\ell\ell',ij}C^{EE}_{\ell',ij}+{\mathcal M}^{--}_{\ell\ell',ij}C^{BB}_{\ell',ij}\right]+N_{\ell}
,\,\,\,\,\,\,\,{\rm and}\,\,\,\,\,\,\,
\widetilde C^{BB}_{\ell,ij}=\left[{\sum_{\ell'}}
{\mathcal M}^{+-}_{\ell\ell',ij}C^{BB}_{\ell',ij}+{\mathcal M}^{-+}_{\ell\ell',ij}C^{EE}_{\ell',ij}\right]+N_{\ell}
\end{eqnarray}
where $N_{\ell}=\sigma^2_e/N_{\rm gal}$ is the shot noise power spectrum (where $N_{\rm gal}$ is the number of galaxies used to compute the power spectrum and $\sigma^2_e$ is the variance of the unlensed ellipticity), and \smash{$C^{EE}_{\ell,ij}$ and $C^{BB}_{\ell',ij}$} are the $EE$ and $BB$ power spectra of shear. If the data is masked or there is a further multiplicative bias then a further mixing of modes would occur to the \smash{$\widetilde C^{EE}_{\ell,ij}$ and $\widetilde C^{BB}_{\ell,ij}$}, as described in \cite{K20}; these additional mixing matrices would affect both  the signal and the noise terms.  

Following \cite{BCT} and Appendix A the calculation of mixing matrices can be written like 
\begin{eqnarray}
\label{matrices2}
    {\mathcal M}^{+\pm}_{\ell\ell',ij}&=&\frac{2\ell'+1}{8\pi}\sum_{\ell''} (2\ell''+1) [1+(-1)^{\ell+\ell'+\ell''}] \left[C^{r_0r_0}_{\ell'',ij} 
    \begin{pmatrix}
    \ell & \ell' & \ell'' \\
    -2 & 2 & 0 \\
    \end{pmatrix}^2    
    +C^{r_4r_4}_{\ell'',ij}
    \begin{pmatrix}
    \ell & \ell' & \ell'' \\
    2 & 2 & -4 \\
    \end{pmatrix}^2
    \pm 2C^{r_0r_4}_{\ell'',ij}
    \begin{pmatrix}
    \ell & \ell' & \ell'' \\
    -2 & 2 & 0 \\
    \end{pmatrix}
    \begin{pmatrix}
    \ell & \ell' & \ell'' \\
    2 & 2 & -4 \\
    \end{pmatrix}\right]\nonumber\\
    {\mathcal M}^{-\pm}_{\ell\ell',ij}&=&\frac{2\ell'+1}{8\pi}\sum_{\ell''} (2\ell''+1) [(-1)^{\ell+\ell'+\ell''}-1]\left[C^{r_0r_0}_{\ell'',ij} 
    \begin{pmatrix}
    \ell & \ell' & \ell'' \\
    -2 & 2 & 0 \\
    \end{pmatrix}^2    
    +C^{r_4r_4}_{\ell'',ij}
    \begin{pmatrix}
    \ell & \ell' & \ell'' \\
    2 & 2 & -4 \\
    \end{pmatrix}^2
    \pm 2C^{r_0r_4}_{\ell'',ij}
    \begin{pmatrix}
    \ell & \ell' & \ell'' \\
    -2 & 2 & 0 \\
    \end{pmatrix}
    \begin{pmatrix}
    \ell & \ell' & \ell'' \\
    2 & 2 & -4 \\
    \end{pmatrix}\right]
\end{eqnarray}
where \smash{$C^{r_Xr_Y}_{\ell,ij}$} is the (cross) power spectrum of $r_X$ and $r_Y$, where $X$ and $Y=\{0,4\}$, and the matrices are Wigner-3$j$ symbols. It is noted that, since the values of $R_{ij}$ are not small compared to the shear, all terms including cross-terms need to be included.

\subsection{Pixel Correction}
An alternative is to correct the observed ellipticity field directly. In this case we label quantities with a subscript $p$ to mean an angular/2-sphere pixels \citep[we use pixelisation of the 2-sphere defined in][]{mw} on the sky e.g. ${\mathbf e}_p$. A correction for the response function then can be constructed as \smash{$\hat{\mathbf \gamma}_p\simeq R^{-1}_p {\mathbf e}_p+{\mathbf \gamma}_p$, 
where $\hat{\mathbf \gamma}_p$} is the estimated shear after correction, which in the case of no noise (${\mathbf e}\rightarrow 0$) is equal to the shear. However, the first term in the correction $R^{-1}_p {\mathbf e}$ represents a local amplification of the shot noise; this is true even if one only applies an average correct by computing the mean of the inverse-response over the sky. This is equivalent of applying a matrix transformation to the unlensed ellipticity where
\begin{eqnarray}
\label{pcorrected}
[\widehat\gamma_1+{\rm i}\widehat\gamma_2]_p&=&[\gamma_1+{\rm i}\gamma_2+
    r'_0(e_1+{\rm i}e_2)+r'_4(e_1-{\rm i}e_2)]_p\nonumber\\
r'_0&=&\frac{1}{2}[(R^{-1}_{11}+R^{-1}_{22})+{\rm i}(R^{-1}_{21}-R^{-1}_{12})]\nonumber\\
r'_4&=&\frac{1}{2}[(R^{-1}_{11}-R^{-1}_{22})+{\rm i}(R^{-1}_{21}+R^{-1}_{12})],
\end{eqnarray}
where $R^{-1}_{ij}$ are elements of the locally-inversed $R$ field on the sphere (i.e. that field which is constructed by computing a pixel-by-pixel inverse of the $R$ field). Therefore when taking the power spectrum the noise-amplification term needs to be corrected. Taking the power spectrum of equation (\ref{pcorrected}) one finds a analogous equation to before where 
\begin{eqnarray}
\label{fullp}
C^{EE}_{\ell,ij}=\widehat C^{EE}_{\ell,ij}-
\left[{\sum_{\ell'}}
{\mathcal N}^{++}_{\ell\ell',ij}N^{EE}_{\ell',ij}+{\mathcal N}^{--}_{\ell\ell',ij}N^{BB}_{\ell',ij}\right]
,\,\,\,\,\,\,\,{\rm and}\,\,\,\,\,\,\,
C^{BB}_{\ell,ij}=\widehat C^{BB}_{\ell,ij}-
\left[{\sum_{\ell'}}
{\mathcal N}^{+-}_{\ell\ell',ij}N^{BB}_{\ell',ij}+{\mathcal N}^{-+}_{\ell\ell',ij}N^{EE}_{\ell',ij}\right]
\end{eqnarray}
where the ${\mathcal N}^{+\pm}_{\ell\ell',ij}$ and ${\mathcal N}^{-\pm}_{\ell\ell',ij}$ are defined in the same way as equation (\ref{matrices2}) except using the  $r'_0$ and $r'_4$ fields defined in equation (\ref{pcorrected}). We note that \smash{$N^{BB}_{\ell,ij}\simeq N^{EE}_{\ell,ij}$} so some terms will cancel.

Thus we find that a local pixel correction requires a correction for the noise amplification effect at the power spectrum level. We note that in doing such a pixel-correction an inverse of the local $R$ matrices is required, which for close-to-singular matrices may lead to an ill-conditioned computational procedure; and since the distribution of $R_{ij}$ values is broad and can cross $R_{ij}=0$ it is possible the matrices may be singular. Hence a free parameter in such an approach is the acceptable condition number allowed, below which a pixel would be masked (thus decreasing the usable area of the survey); we define the conditional number\footnote{See \url{https://numpy.org/doc/stable/reference/generated/numpy.linalg.cond.html}.} for $x$ as the norm of $x$ multiplied by the norm of the inverse of $x$, i.e. $c_R=|x|/|x^{-1}|$. To explain further: in the pixel-correction case one needs to invert the $R_{ij}$ matrices locally, the matrices might not be numerically invertible everywhere, we use the condition number to determine what is invertible, we consider a pixel unobserved if its matrix is not invertible. In the next Section we create simple simulations to test the forward-modelling approach, the pixel-correction approach, and an approach of taking the mean response over the sky. 

\section{Tests on Simulations}
\label{S:Results}

We model $\gamma(\mathbf{\Omega})$ as a Gaussian random field \citep[generateed using the {\tt massmappy} code][]{wallis}, assuming a DES Year 1 cosmology \citep{des1,des2,des3} to compute the EE cosmic shear power spectrum. We assume the the Limber \citep{1953ApJ...117..134L,2017MNRAS.469.2737K,2017JCAP...05..014L}, reduced shear \citep{ad2}, flat-Universe \citep{2018PhRvD..98b3522T}, flat-sky \citep{10.1046/j.1365-8711.1998.02054.x} and prefactor-unity \citep{2017MNRAS.469.2737K} approximations. Therefore the EE power spectrum is given by:
\begin{equation}
    \label{eq:cltheory}
    C^{EE}_{\ell} = \int_0^{\chi_{\rm H}} {\rm d}\chi \frac{q^2(\chi)}{\chi^2} P_{\delta} \left(\frac{\ell + 1/2}{\chi}, \chi\right),\,\,\,\,\,\,\,{\rm where}\,\,\,\,\,\,\,
    q(\chi) = \frac{3}{2}\Omega_{\rm M} \frac{H^2_0}{c^2} \frac{\chi}{a(\chi)} \int^{\chi_{\rm H}}_\chi {\rm d}\chi'\, n(\chi')\, \frac{\chi'-\chi}{\chi};
\end{equation}
where $P_{\delta}$ is the power spectrum of matter overdensities that we calculate using \texttt{CAMB} \citep{cambcite} \citep[we include the corrections from][for the non-linear corrections]{2015MNRAS.454.1958M}. $H_0$ is the Hubble constant, $\chi$ and $\chi_{\rm H}$ are the comoving distance and comoving distance to the horizon respectively \citep[calculated using the \texttt{astropy} package][]{astropy1, astropy2}, $a$ is the scale factor of the Universe, $\Omega_{\rm M}$ is the dimensionless total matter density of the Universe, and $c$ is the speed of light in a vacuum. $n(\chi)$ is the galaxy distribution function of the survey \citep[where we use the photometric DES Year 1 galaxy distribution.][]{des2}, and we assume only a single tomographic bin. We assume that $EB$ and $BB$ shear power spectra are zero.
\begin{figure*}
\centering
\includegraphics[width=0.48\columnwidth]{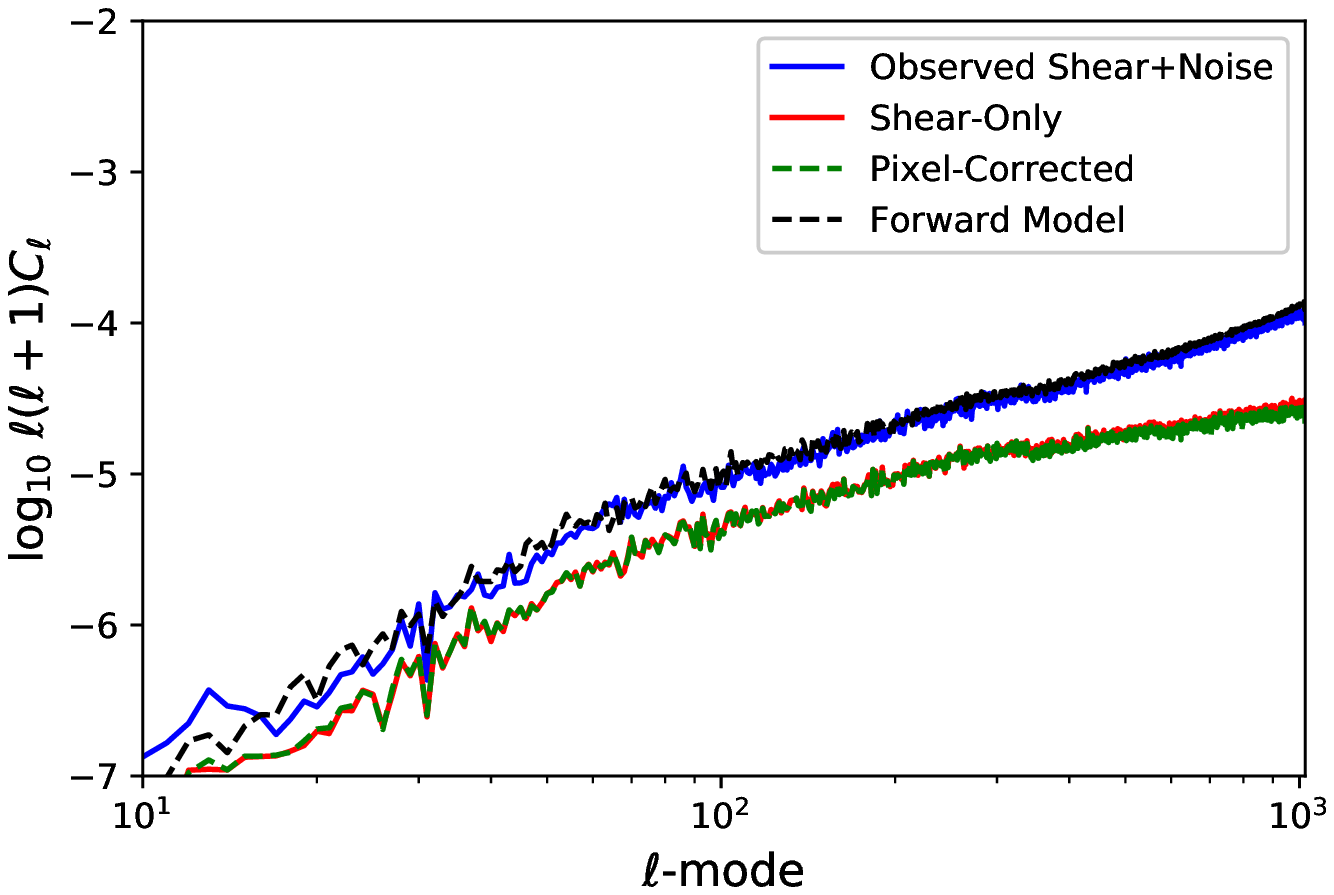}
\includegraphics[width=0.48\columnwidth]{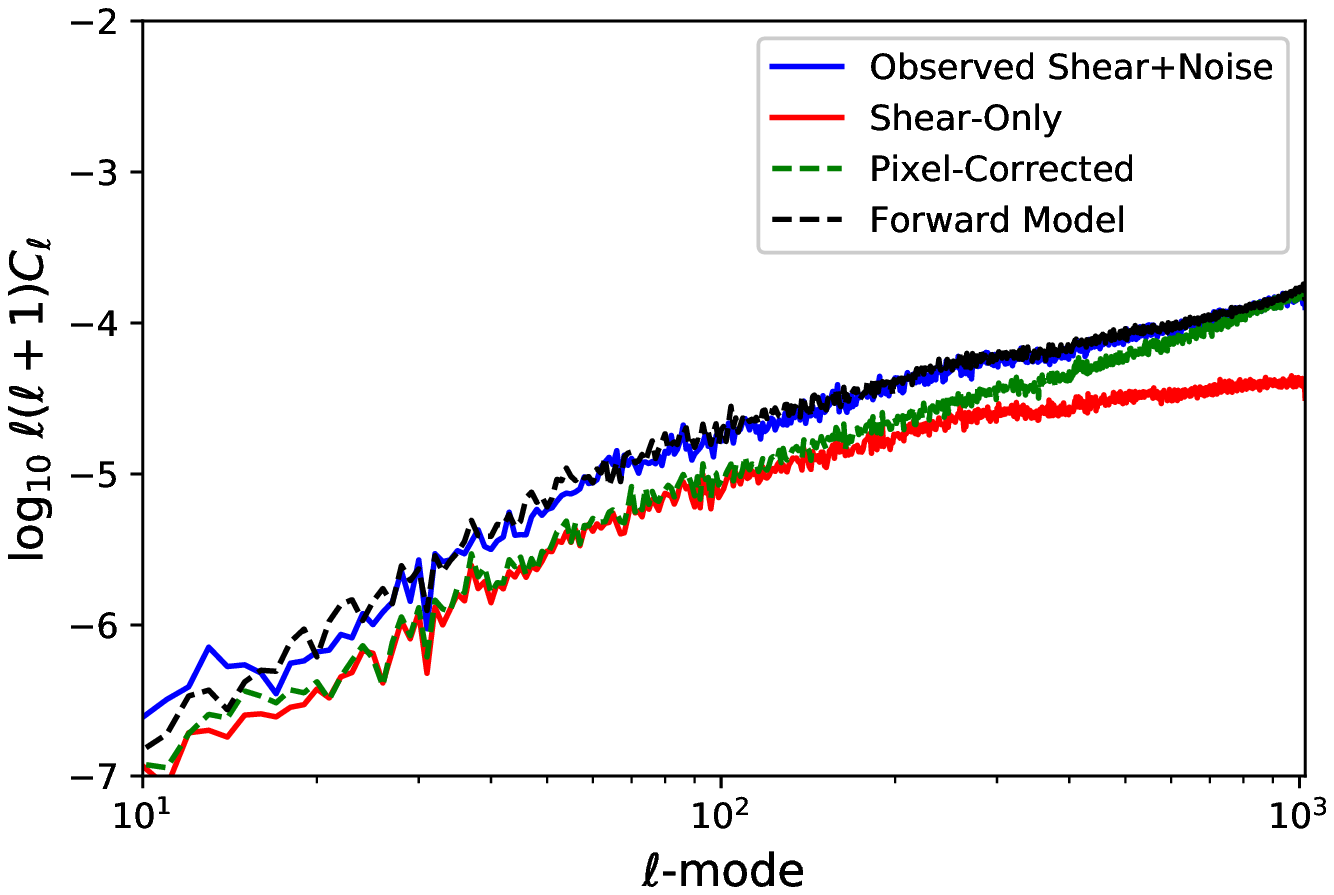}\\
\caption{We show the observed power spectrum, that includes the effects of the response matrix and noise (blue lines); and shear-only power spectrum that does not include noise or the response matrix terms (red lines). We then compare this the pixel-correction approach (equation \ref{pcorrected}, green lines) that should reproduce the shear-only power spectrum, and the forward modelling approach (equation \ref{full}; blue lines) that should produce the observed power spectrum. The left-hand plot is for a condition number limit of $c_R\leq 10$, and the right-hand for $c_R\leq 100$ in the pixel-corrected approach.} 
\label{Test}
\end{figure*}

We also make a mask that removes data within $20^{\circ}$ in both the galactic and ecliptic planes; and also $20\%$ of pixels at random, to represent an all-sky mask with random patches removed, resulting in a total observed sky fraction of $f_{\rm sky}=0.4$. Finally we assume $\sigma_e=0.3$, and $N_{\rm gal}=30f_{\rm sky}3600(4\pi[180/\pi]^2)$ for the shot noise modelling. This is is similar to upcoming Stage-IV surveys \citep{2006astro.ph..9591A}. We then model the response matrix elements as Gaussian random fields with a white-noise power spectrum with mean $\mu=1$ and a standard deviation of $\sigma=0.6$ for all $R_{ij}$; which approximates the amplitude observed in real shape measurement methods \citep{metacal}. This is meant as the simplest test of the approaches outlined in this paper, in future much more realistic values should be used that also correlate with galaxies, instrument and telescope properties. 

In Figure \ref{Test} we show the observed power spectrum, that includes the effects of the response matrix and noise (blue lines); and shear-only power spectrum that does not include noise or the response matrix terms (red lines). We then compare this the pixel-correction approach (equation \ref{pcorrected}) that should reproduce the shear-only power spectrum, and the forward modelling approach (equation \ref{full}) that should produce the observed power spectrum. We test the pixel-correction approach for two cases where the condition number of the response matrices is limited to $c_R \leq 10$ and $c_R\leq 100$; where $c_R$ is the condition number of the local response matrices, defined using the {\tt numpy} function {\tt numpy.linalg.cond}$^{1}$. We find that both approaches reproduce the expected results, however in the pixel-correction approach setting a condition number limit on the matrices that is too large can lead to an amplification of the noise term in equation (\ref{pcorrected}) and hence a deviation from the shear-only power spectrum. Thus there is a trade-off in the pixel-correction approach between accuracy (a tight limit on the condition number) and area coverage (i.e. pixels not included because they are excluded by the limit). In our simple simulation we find a 20\% change in area caused by going from $c_R\leq 100$ to $c_R\leq 10$. Such a trade-off is not required in the forward modelling approach. In the case of tomographic binning the shot-noise amplification will be larger because there will be fewer galaxies in each pixel.

Finally a different approach could instead use a global correction, by dividing all observed ellipticities by the mean of the response function over the sky, hoping to derive the true shear power spectrum. In this case, since in our simulations the mean is unity, this would result in an error between the inferred shear power spectrum and the true shear power spectrum that was equivalent to the difference between the uncorrected case (blue lines) and the true case (red lines). Clearly in such an approach there would be a large bias. 

\section{Conclusions}
\label{Conclusions}
In this paper we show how response functions propagate into cosmic shear power spectra computed from pixelised maps, and we investigate pixel-correction and forward-modelling approaches using simulations. In the pixel-correction approach there is a free parameter that is the tolerated condition number of the local response matrices -- if this is too large then this can cause an amplification of the shot noise power spectrum, if too small it can lead to a loss of area. In contrast forward-modelling the power spectrum avoids this choice. Forward-modelling involves measuring the response of each galaxy as a function of local conditions at the map level and the propagation of angular variation through to the power spectrum. In more complex approaches one could also draw from the probability distribution function of the response-function map to propagate uncertainties in these measurements into the power spectrum. Alternatively one could infer the power spectrum directly, rather than via a map, but in this case the angular variation of the response function would still need to be accounted for.

\vspace{-0.5cm}
\acknowledgements
\noindent{\scriptsize \emph{Acknowledgements:} We thank the developers of {\tt SSHT}, {\tt massmappy}, {\tt astropy} and {\tt CAMB}. NT is supported by UK Space Agency grants ST/W002574/1 and ST/X00208X/1.} 
\vspace{-0.3cm}
\bibliographystyle{mnras}
\bibliography{sample.bib}

\section*{Appendix A: Product of Spherical Spin-Weight Functions on the Sphere}

Here we follow \cite{NISTDLMF} Section 34.3 and \cite{2019arXiv190409973T}. Let $f({\mathbf{\Omega}})$ and $g({\mathbf{\Omega}})$ be spin-weighted spherical functions with respective spins
$s_f$ and $s_g$. The product $h({\mathbf{\Omega}}) = f({\mathbf{\Omega}})g({\mathbf{\Omega}})$ has spin weight $s = s_f + s_g$. The spherical harmonic transform of $h({\mathbf{\Omega}})$ can be written in terms of spin-weighted spherical harmonic functions $_{s}Y_{\ell m}(\mathbf{\Omega})$
\begin{equation}
    h_{\ell m}=\sum_{\ell_f m_f}\sum_{\ell_g m_g}f_{\ell_f m_f}g_{\ell_g m_g}\int {\rm d}{\mathbf{\Omega}}\,\,_{s_f}Y_{\ell_f m_f}(\mathbf{\Omega})_{s_g}Y_{\ell_g m_g}(\mathbf{\Omega})_{s}Y^*_{\ell m}(\mathbf{\Omega}),
\end{equation}
where $f_{\ell_f m_f}$ and $g_{\ell_f m_f}$ are the spherical harmonic transforms of $f$ and $g$; $\ell$ and $m$ are wavenumbers. The integral can be rewritten so that 
\begin{equation}
    h_{\ell m}=\sum_{\ell_f m_f}\sum_{\ell_g m_g}(-1)^{s+m}f_{\ell_f m_f}g_{\ell_g m_g}\left(\frac{(2\ell_f+1)(2\ell_g+1)(2\ell+1)}{4\pi}\right)^{1/2}
    \begin{pmatrix}
    \ell_f & \ell_g & \ell \\
    -s_f & -s_g & s \\
    \end{pmatrix}
    \begin{pmatrix}
    \ell_f & \ell_g & \ell \\
    m_f & m_g & -m \\
    \end{pmatrix},
\end{equation}
this is the spherical harmonic convolution theorem for products of functions, where the matrices are Wigner-$3j$ symbols. To compute the effect of the convolution on the angular
power spectrum we consider the products
$h = fg$ and $h' = f'g'$. In general it is not possible to express the cross-power spectrum of $h$ and $h'$, $C^{hh'}_{\ell}$ in terms of combinations of the power spectrum of the individual fields. However if $f$ and $f'$ are jointly homogeneous spherical random fields, independent of $g$ and $g'$, we can write down an expression for the expectation of the angular power spectrum $\langle C^{hh'}_{\ell}\rangle$; since the differently-numbered modes of any jointly homogeneous random fields are uncorrelated i.e. \smash{$\langle f_{\ell m}(f'_{\ell' m'})^*\rangle=\delta^K_{\ell\ell'}\delta^K_{mm'}\langle C^{ff'}_{\ell}\rangle$}, where $\delta^K_{xx'}$ is a Kronecker delta. Using this, and the orthogonality relation for Wigner-$3j$ symbols we can write down the final result as the convolution
\begin{eqnarray}
\label{mixm}
\langle C^{hh'}_{\ell}\rangle=\sum_{\ell_f\ell_g} (-1)^{s+s'}\frac{(2\ell_f+1)(2\ell_g+1)}{4\pi}
\langle C^{ff'}_{\ell_f}\rangle C^{gg'}_{\ell_g}
    \begin{pmatrix}
    \ell_f & \ell_g & \ell \\
    -s_f & -s_g & s \\
    \end{pmatrix}
        \begin{pmatrix}
    \ell_f & \ell_g & \ell \\
    -s'_f & -s'_g & s' \\
    \end{pmatrix}.
\end{eqnarray}
This can be written as a
matrix multiplication \smash{$\langle C^{hh'}_{\ell}\rangle=\sum_{\ell_f}{\mathcal M}^{gg'}_{\ell\ell_f}\langle C^{ff'}_{\ell_f}\rangle$} where ${\mathcal M}^{gg'}_{\ell\ell_f}$ can be inferred from equation (\ref{mixm}). 
\label{lastpage}
\end{document}